\documentclass[11pt,a4paper]{article}

\usepackage{SGpreprint}

\nologohead{Frank Schweitzer: Multi-Agent Approach to the Self-Organization
  of Networks \\ 
\emph{in}: F. 
Reed-Tsochas, N. F. Johnson,  J. Efstathiou: \textsl{Understanding
and Managing Complex Agent-Based Dynamical Networks}, 
Singapore: World Scientific (2007) \\
see \url{http://www.sg.ethz.ch} for more information
} 

\usepackage{epsfig}

\begin{document}

\begin{center}

   \textbf{\Large Multi-Agent Approach to the Self-Organization of
     Networks}\footnote{This paper was finished December 2005}\\[5mm]
   \textbf{\large Frank Schweitzer}

       Chair of Systems Design, ETH Zurich, Kreuzplatz 5, 
       8032 Zurich, Switzerland \\
       \url{fschweitzer@ethz.ch}
\end{center}

\newcommand{\mean}[1]{\left\langle #1 \right\rangle}
\newcommand{\abs}[1]{\left| #1 \right|}
\newcommand{\bbox}[1]{\mbox{\boldmath $#1$}}
\newcommand{\eps}{\varepsilon}
\renewcommand{\epsilon}{\varepsilon}
\renewcommand{\footnoterule}{\vspace{0.5cm}%
\rule{2.5in}{0.4pt} \vspace{0.3cm}}
\newcounter{myfn}[page]
\renewcommand{\thefootnote}{\fnsymbol{footnote}}
\newcommand{\myfootnote}[1]{\setcounter{footnote}{\value{myfn}}%
\footnote[2]{#1}\stepcounter{myfn}}
\newcommand{\fn}[1]{\myfootnote{ #1}}


\section{Self-Organization of Networks}
  
\emph{Self-organization} denotes ``the process by which individual
subunits achieve, through their cooperative interactions, states
characterized by new, emergent properties transcending the properties of
their constitutive parts.'' \citep{biebrep-95} Whether these
emergent properties occur or not depends of course not only on the
properties of the subsystems and their interactions, but also on suitable
external conditions, such as global boundary conditions, the in/outflux
of resources (free energy, matter, or information).  A description that
tries to include these conditions is given by the following heuristic
definition: ``Self-organization is defined as spontaneous formation,
evolution and differentiation of complex order structures forming in
non-linear dynamic systems by way of feedback mechanisms involving the
elements of the systems, when these systems have passed a critical
distance from the statical equilibrium as a result of the influx of
unspecific energy, matter or information.''  \citep{sfb-es-96}

In this sense, the self-organized structure formation can be considered
as the opposite of a hierarchical design of structures which basically
proceeds {\em from top down to bottom}: here, structures are {\em
  originated} bottom up, leading to an emerging \emph{hierarchy}, where
the structures of the ``higher'' level appear as a new quality of the
system \citep{haken-78,darley-94}.  For the prediction of these global
qualities from local interactions fundamental limitations exist which are
discussed e.g. in chaos theory.  Moreover, stochastic fluctuations also
give unlikely events a certain chance to occur, which in turn affects the
real history of the system.  This means, the properties of complex
systems cannot be determined by a hierarchy of conditions, the system
creates its complexity in the course of evolution with respect to its
global constraints. Considering that also the boundary conditions may
evolve and new degrees of freedom appear, \emph{coevolutionary processes}
become important, and the evolution may occur on a qualitatively new
level.

The self-organization of \emph{network structures}, i.e.  the
\emph{emergence} of links between a set of nodes is of crucial importance
in many different fields. In electronic engineering, for instance, one is
interested in the \emph{self-assembling} and \emph{self-repairing} of
electronic circuits \citep{wang-01,cui-lieber-01,mange-stauffer-94},
while in biology models for the self-wiring of \emph{neuronal networks}
are investigated \citep{segev-physlett-98, segev-00}.  On the social
level, the self-organization of human trail networks between different
destinations is a similar problem \citep{helbing-fs-et-97}.  Also the
establishment of connections on demand in telecommunication or logistics
is related to the problem discussed here.

In general, the self-organized formation of a network is rather different
from drawing lines between a set of nodes. Often no \emph{a priori}
information about the network structure is provided, i.e.  the network
has to self-organize itself not only regarding the links but also
regarding the nodes. This is the case for instance if the nodes to be
linked to the network are ``unknown'' in the sense, that they
\emph{first} have to be \emph{discovered} and only \emph{then} can be
\emph{connected}. 

A common biological example is the formation of a trail system in ants
(cf also Secs. 6, 7) to connect a nest to a set of food sources that first
have to be found \citep{crist-haefner-94, fs-lao-family-97}. Such
networks are known to be rather flexible and adaptive.  After food
sources are exhausted, they are ``disconnected'' from the existing
network, because they are no longer visited and the respective trail is
no longer maintained, but newly found food sources can be linked to the
existing network as well.
Hence, \emph{adaptivity} is a desirable feature of self-organized
networks.  This means that new nodes can be linked to the existing
network or linked nodes can be disconnected from the network if this is
required, e.g. by the change of some external conditions.  

Noteworthy, such a behavior should be not governed by a ``supervisor'' or
``dispatcher'', it should rather result from the adaptive capabilities of
the network itself. In the self-wiring of neural structures, for example,
a neuron grows from the retina of the eye towards the optic tectum (or
superior colliculus) of the brain, without ``knowing'' from the outset
about its destination node in the brain. Hence it has to navigate through
an unknown environment in order to detect and to reach the appropriate
area. It is known that gradients of different chemical cues play a
considerable role in this navigation process. They provide a kind of
\emph{positional information} for the navigation of the growth cones
\citep{gierer-83}. But in the very beginning, this positional information
has to be generated interactively, and only in later stages may lead to
established pathways.

This points to the questions that shall be discussed in this chapter: Is it
possible to link a set of nodes without using preexisting positional
information or any kind of long-range attraction of the nodes? Can the
process of generating positional information, i.e. the detection of
``unknown'' nodes and the estabishment of chemical gradients, \emph{and}
the process of network formation, i.e. the establishment of links between
nodes, occur in parallel, on a comparable time scale, as a process of
co-evolution?

In the following, we will discuss a model where the generation of
relevant information for establishing the links between nodes results
from the interaction of many \emph{agents}, i.e.  subunits of the system
that are capable of performing some activities.  Their collective
interaction is based on (indirect) communication, which also includes
memory effects and the dissemination of information in the system. The
relevant (``pragmatic'') information that leads to the establishment of
the links then emerges from an evolutionary interplay of selection and
reamplification. In the next section, we will introduce the general
framework for modeling such processes, while in Secs. 3-7 two different
applications of the model will demonstrate its suitability.


\section{Brownian Agents}

Recently, different computer architectures in distributed artificial
intelligence have been developed to simulate the collective behavior of
interacting agents (cf. for instance the \textsc{Swarm}
project at \url{http://www.swarm.org/}). However, due to their rather
complex simulation facilities many of the currently available simulation
tools lack the possibility to investigate systematically and in depth the
influence of specific interactions and parameters. Instead of
incorporating only as much detail as is \emph{necessary} to produce a
certain emergent behavior, they put in as much detail \emph{as possible},
and thus reduce the chance to understand \emph{how} emergent behavior
occurs and \emph{what} it depends on.

Therefore, it would be feasible to have multi-agent systems (MAS) that
can be also investigated by means of \emph{analytical methods} (from
statistical physics or mathematics) -- in addition to their computational
suitability. The concept of \emph{Brownian agents} \citep{fsbook03} is
one of the possible approaches to serve for this purpose. It denotes a
particular class of agents that combines features of reactive and
reflexive agent concepts.  A reflexive agent has an (internal) model or
at least some knowledge about its environment that allows its to draw
conclusions about some certain actions in a given situation.  The
reactive agent, on the other hand, simply ``reacts'' to signals from the
environment without referring to internal knowledge.  This
signal-response type of action implies the same reaction to the same
signal. From this perspective, the reactive agent is a rather
\emph{simplex} or \emph{minimalistic} agent, while the reflexive agent is
a rather \emph{complex} agent.

The \emph{Brownian agent} \citep{fsbook03} is more advanced than the
reactive agent but not yet of the internal complexity of the reflexive
agent. It can be described by a set of state variables $u_{i}^{(k)}$.
The index $i=1,...,N$ refers to the individual agent $i$, while $k$
indicates the different variables.  These could be either \emph{external}
variables that can be observed from the outside, or \emph{internal
  degrees of freedom} that can only be indirectly concluded from
observable actions.  Important external variables are
$u^{(1)}_{i}=\bbox{r}_{i}$, which denotes the \emph{space coordinate}
(mostly a vector in the two-dimensional physical space), or
$u^{(2)}_{i}=\bbox{v}_{i}$, which is the individual \emph{velocity} in
the case of a moving agent. Both are assumed as \emph{continuous}
variables.

The internal degrees of freedom, on the other hand, cannot be directly
observed. They could be continuous or discrete variables. For instance,
the state variable $u^{(3)}_{i}=\theta_{i}\in \{-1,0,+1\}$ may
distinguish between three different responses to certain environmental
conditions or to incoming information.  For example, agents with
$\theta=-1$ may not be affected by a particular signal, while agents with
$\theta=+1$ may respond to it.  An important continuous state variable in
the context of Brownian agents is the \emph{internal energy depot}
$u^{(4)}_{i}=e_{i}$, which determines whether agent $i$ may perform a
certain action or not.  This includes the assumption that all actions --
be it active motion or communication or environmental changes -- need to
use ``energy''. In general, this term describes not just the physical
free energy that is dissipated e.g.  during active motion, it intends to
cover also other resources needed to perform a certain action.
   
Noteworthy, the different (external or internal) state variables can
change in the course of time, either due to impacts from the surrounding,
or due to an internal dynamics. Thus, in a most general way, we may
express the dynamics of the different state variables as follows:
\begin{equation}
  \label{u-ik-d}
  \frac{d\,u_{i}^{(k)}}{dt}=f_{i}^{(k)}+\mathcal{F}^{\mathrm{stoch}}_{i}
\end{equation}
For the Brownian agents, it is assumed that the causes for the
temporal change of $u_{i}$ may be described as a
\emph{superposition} of \emph{deterministic} and \emph{stochastic}
influences, imposed on agent $i$.  This picks up the ingenious idea first
used by Langevin in early 1900 to describe the motion of
\emph{Brownian particles} -- and is basically the reason why this agent
concept is denoted as \emph{Brownian} agent.  A Brownian particle moves due
to the impacts of the surrounding molecules whose motion however can be
observed only on a much smaller time and length scale compared to the
motion of the Brownian particle.  Thus, Langevin invented the idea
to sum up all these impacts in a stochastic force with certain
statistical properties.

For the Brownian agent, we will exploit Langevins idea in a
similar manner, i.e. we will sum up influences which may exist on a
microscopic level, but are not observable on the time and length scale of
the Brownian agent, in a stochastic term $\mathcal{F}_{i}^{\mathrm{stoch}}$,
while all those influences that can be directly specified on these time
and length scales are summed up in a \emph{deterministic} term
$f_{i}^{(k)}$.  Such a distinction basically defines the level of
coarse-grained description for the multi-agent system.  The ``cut'' may
prevent us from considering too much ``microscopic'' details of the MAS,
while focussing on particular levels of description.  The summed up
stochastic influences might result from a more fine-grained deterministic
description -- but instead of taking this into detailed account, just
some specific statistical (gross) properties are considered on the
coarse-grained level.  Notheworthy, the strength of the stochastic
influences may also vary for different agents and may thus depend on
local parameters or internal
degrees of freedom. 
  
The \emph{deterministic} part $f_{i}^{(k)}$ contains all specified
influences that cause changes of the state variable $u_{i}^{(k)}$. This
could be non-linear interactions with other agents $j\in N$, external
conditions - such as forces resulting from external potentials, or the
in/outflux of resources etc. or even an \emph{eigendynamics} of the
system that does not depend on the action of the agents. In the example
of an ecosystem this eigendynamics may describe day/night or seasonal
cycles, or the agent-independent diffusion of resources within the
system.

\section{Basic Agent Model of Network Formation}

In order to set up a Brownian multi-agent model for the formation of
networks we need to specify (i) the relevant state variables
$u_{i}^{(k)}$ and (ii) the dynamics for changing them, i.e.
$\dot{u}_{i}^{(k)}$. For the application considered, we may assume that
each Brownian agent is characterized by three state variables: spatial
position $\bbox{r}_{i}$, (discrete) internal state $\theta_{i}$ and
(continuous) internal energy depot $e_{i}$.  For the change of the
agent's position we may assume an overdamped Langevin equation:
\begin{equation}
\frac{d\bbox{r}_{i}}{dt}= f_{i}
+ \sqrt{2\eps_{i}}\,\xi_{i}(t)
\label{langev-red2}
\end{equation} 
The second term denotes the stochastic influences, where $\xi_{i}(t)$ is
white noise with $\mean{\xi_{i}(t)}=0$ and
$\mean{\xi_{i}(t)\,\xi_{j}(t')} = \delta_{ij}\,\delta(t-t')$.  The
strength of the stochastic force $\eps_{i}$ could be in general an
individual parameter to weight the stochastic influences, this way it can
for example measure the individual {\em sensitivity} $\eta_{i}\propto
1/\eps_{i}$ of the agent. The deterministic influences in the first term,
$f_{i}$, are in the considered case assumed to result from the the
(direct or indirect) interaction between the agents, as specified below.

\begin{figure}[ht]
\centerline{
\includegraphics[width=6cm,height=6cm]{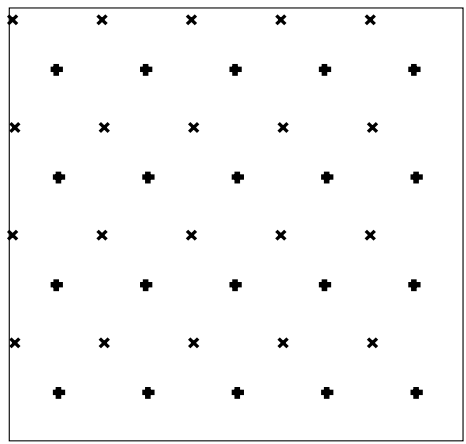}
}
\caption[p-raut0]{
  Example of a regular distribution of 40 nodes on a lattice of size
  $A=100 \times 100$. For the computer simulations, periodic boundary
  conditions have been used. $z_{+}=20$, $z_{-}=20$. \{x\} indicates
  nodes with a potential $V_{j}=-1$, \{+\} indicates nodes with a
  potential $V_{j}=+1$. \label{raut0}  \citep{fs-tilch-02-a}}
\end{figure}

The internal degree of freedom $\theta_i(t)$ should have one of the
following values: $\theta_{i} \in \{0,-1,+1\}$.  Initially,
$\theta_i(t_0)=0$ holds for every agent. The variable $\theta_i$ can be
changed in the course of time by an interaction between the moving agents
and the nodes. To be specific, we consider a two-dimensional surface,
where a number of $j=1,...,z$ nodes are located at the positions
$\bbox{r}^{z}_{j}$ (cf.  Fig. \ref{raut0}). A number of $z_+$ nodes
should be characterized by a positive potential, $V_j=+1$, while
$z_-=z-z_+$ nodes have a negative potential, $V_j=-1$.  We note
explicitely, that the nodes do \emph{not} have any \emph{long-range
  effect} on the agents, such as attraction or repulsion. Their effect is
restricted to their location, $\bbox{r}^{z}_{j}$.

It is the (twofold) task of the Brownian agents, first to \emph{discover}
the nodes and then to \emph{link} nodes with an opposite potential, this
way forming a self-organized network between the set of nodes. If an
agent hits one of the nodes, its internal degree of freedom is changed
due to the following equation:
\begin{equation}
\Delta \theta_i(t)= 
 \sum_{j=1}^z (V_j-\theta_i)\; \int_{A} \frac{1}{A}
\,\delta\Big(\bbox{r}^{z}_{j}-\bbox{r}_{i}(t)\Big) \,d\bbox{r} 
\label{change}
\end{equation}
The delta function is equal to $1$ only for
$\bbox{r}^{z}_{j}=\bbox{r}_{i}$ and zero otherwise. So, eq.
(\ref{change}) indicates, that an agent can change its internal state,
$\theta_i$, only if it hits one of the nodes. Then it takes over the
value of the potential of the respective node, $V_j$, which means
$\theta_i$ remains constant if $V_j=\theta_i$, and $\theta_i \rightarrow
V_j$, if $V_j \neq \theta_i$.  We note that the probability for a
(pointlike) agent to hit a (pointlike) node is almost vanishing. However,
the computer simulations discussed in the following section are carried
out on a discrete lattice, so the agent and the node both have a finite
extension, in which case Eq. \ref{change} makes sense.

Eventually, we have to describe the dynamics of the internal energy
depot, $e_{i}$, of the agent. In this specific application, it is assumed
that every time $t^{i}_{n+}$, $t^{i}_{n-}$ when Brownian agent $i$ hits
one of the nodes $n+$ or $n-$ its internal energy depot $e_{i}$ is filled
up to a maximum value $e_{\mathrm{max}}$. Then, the agent is in an active
state that allows for certain actions.  In our case, the agent is able to
produce a chemical, either component $(-1)$ or $(+1)$.  Which \emph{kind}
of chemical is produced depends on the actual value of the internal
parameter $\theta_{i}(t)$ of the agent as specified below, whereas the
\emph{amount} of chemical produced depends on the actual value of the
internal energy depot $e_{i}(t)$. For simplicity, we assume for the
amount of the production rate
\begin{equation}
  \label{eq:se}
  s_{i} \propto e_{i} \;\to\; s_{i}(\theta_{i},t)=\beta\,e_{i}(t)
\end{equation}
The internal energy depot can only be charged when hitting one of the
nodes and can be filled only to a maximum $e_{\mathrm{max}}$. On the
other hand, it is decreased by the production of the chemical.  Hence, in
the general balance equation 
\begin{equation}
  \label{eq:edot}
\frac{d\, e_{i}}{dt} = q_{i}(\bbox{r}_{i},t) 
- p_{i}(\bbox{r}_{i},t) 
\end{equation}
the space and time dependent influx of energy into the internal depot is
given by:
\begin{equation}
  \label{eq:qrt}
  q_{i}(\bbox{r}_{i},t)= \big(e_{\mathrm{max}} -e_{i}(t)\big)
\,\delta(\bbox{r}_{j}^{z}-\bbox{r}_{i})\,\delta(t-t^{i}_{n\pm})
\end{equation}
wheras the ``outflux'' of energy from the agent's depot,
$p_{i}(\bbox{r}_{i},t)$, reads accordingly:
\begin{equation}
  \label{pi}
  p_{i}(\bbox{r}_{i},t)=s_{i}(\theta_{i},t)
\end{equation}
We note that in previous investigations
\citep{fs-eb-tilch-98-let,eb-fs-tilch-99,eb-fs-03} an additional term for the
active
(accelerated) motion of the agent was included in Eq. \ref{pi}. Further, 
a dissipative loss, $-ce_{i}$, of the internal energy depot was
considered. Both terms are neglected here.

Considering Eq. (\ref{eq:se}), we can deduce from Eq.~(\ref{eq:edot})
the time-dependence of the production rate as $s_{i}(\theta_{i},t) \sim
\exp{\{-\beta t\}}$. The dependence on $\theta_{i}$, however, involves
specific assumptions of our network model:
\begin{equation}
  \begin{aligned}
s_i(\theta_i,t)=s_{\mathrm{max}}\,\frac{\theta_i}{2}\Big[&(1+\theta_i)\,
\exp\{-\beta\,(t-t_{n+}^{i})\}\\  
-\,&(1-\theta_i)\,\exp\{-\beta\,(t-t_{n-}^{i})\}\Big]
\label{prod}
\end{aligned}
\end{equation}
Eq.  (\ref{prod}) means that the agent is not active, as long as
$\theta_i=0$, which means before it hits one of the nodes the first time.
After that event, the agent begins to produce either component $(+1)$ if
$\theta_i=+1$, or component $(-1)$ if $\theta_i=-1$, at a rate that
exponentially decreases with time. The maximum production rate
$s_{\mathrm{max}}$ depends on the maximum amount $e_{\mathrm{max}}$,
after the internal energy depot was charged at one of the nodes.

The spatio-temporal concentration of the chemicals shall be described by
a \emph{chemical field} $h_{\theta}(\bbox{r},t)$ consisting of the two
components $(+1)$ or $(-1)$, which obeys the following dynamics:
\begin{equation}
\frac{\partial h_{\theta}(\bbox{r},t)}{\partial t}=-k_{h}\,
h_{\theta}(\bbox{r},t) + \sum_{i=1}^{N} 
s_i(\theta_i,t)\;\delta_{\theta;\theta_{i}}\;
\delta\Big(\bbox{r}-\bbox{r}_i(t)\Big) 
\label{h-net-nd}
\end{equation}
The first term describes the exponential decay of the existing
concentration due to spontaneous decomposition of the chemical, where
$k_{h}$ is the decomposition rate. The second term denotes the
production of the field by the agents.  Here,
$\delta_{\theta;\theta_{i}}$ means the Kronecker Delta used for discrete
variables, indicating that the agents only contribute to the field
component that matches their internal parameter $\theta_{i}$. The Delta
function $\delta(\bbox{r}-\bbox{r}_i(t))$ means that the agents
contribute to the field only \emph{locally}, at their current position,
$\bbox{r}_{i}$.  Diffusion of the chemical substances is not
considered here.

As a last step, we have to specify how the existence of the two different
chemicals affect the motion of a Brownian agent, described by
Eq.~(\ref{langev-red2}). We assume that the interaction term $f_{i}$
results from an \emph{indirect communication} among the agents. By
producing one out of two chemicals at its current position, each agent
generates \emph{local information} that is stored outside the agent, i.e.
the environment acts as a kind of external memory. This information
should be accessible to other agents in the vicinity. Specifically, we
assume that the agent is sensitive to local \emph{gradients} in these two
chemical components dependent on its current internal state $\theta_{i}$:
\begin{equation}
f_{i}= \alpha\,\frac{\theta_i}{2}\!\left[ (1+\theta_i)\left.\frac{\partial
        h_{-1}(\bbox{r},t)}{\partial \bbox{r}}\right|_{r_{i}}-
      (1-\theta_i)\left.\frac{\partial h_{+1}(\bbox{r},t)}{\partial
        \bbox{r}}\right|_{r_{i}}\right]
\label{dh-eff}
\end{equation}
Here, $\alpha$ is a dimensional constant only, but it could also act as
an individual response parameter weigthing the importance of the
information received \citep{fs-97-agent}.

This way, by producing and responding to two different local information
in the system, the Brownian agents communicate with each other. The
chemical field plays the role of an order parameter. It does not exist
from the outset, but is generated by the agents. Once it is established,
it reduces the ``freedom'' of the moving agents by guiding their motion
using the information generated by other agents. 

\begin{figure}[htbp]
\centerline{
\includegraphics[width=7.cm]{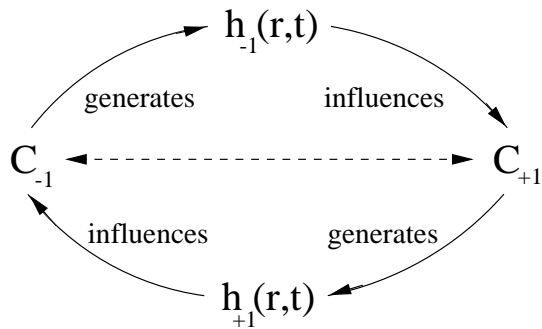}
}
\caption[]{
  Circular causation between the  Brownian agents in the
  different internal states, $C_{-1}$, $C_{+1}$, and the two-component
  chemical field, $h_{\theta}(\bbox{r},t)$ \label{caus2a}}
\end{figure}
Fig. \ref{caus2a} summarizes the non-linear feedback between the
two-component chemical field and the agents, as given by the Eqs.
(\ref{prod}), (\ref{dh-eff}). Our model assumes, that agents with an
internal state $\theta_i=0$ do not contribute to the field and are not
affected by the field. They simply move like Brownian particles. Agents
with an internal state $\theta_i=+1$ contribute to the field by producing
component $+1$, while they are affected by the part of the field that is
determined by component $-1$. On the other hand, agents with an internal
state $\theta_i=-1$ contribute to the field by producing component $-1$
and are affected by the part of the field, which is determined by
component $+1$.  Moreover, if the agent hits one
of the nodes, the internal state can be switched due to Eq.
(\ref{change}).  Hence, the agent begins to produce a different chemical,
while affected by the opposite potential. Precisely, at one time the
agent does \emph{not} respond to the gradient of the same field, which it
contributes to via producing a chemical.

As the result of this nonlinear feedback between the Brownian agents and
the chemical field generated by them, we can observe the selforganized
formation of a network shown in the following section.

\section{Computer Simulations of Network Formation}
\label{sec:sim}

For the computer simulations, a triangular lattice with periodic boundary
conditions has been used. 
The agents start initially at random positions. For the evolution of the
network, we evaluate the sum $\hat{h}(\bbox{r},t)$ of the two field
components generated by the agents.  For the plots, however, we have to
match these values with a \emph{grey scale} of 256 values, which is
defined as follows:
\begin{eqnarray}
  \label{grey}
c(\bbox{r},t) &=& 255 \left[1- \log\left(1 + 9\,
\frac{\hat{h}(\bbox{r},t)-\hat{h}_{\mathrm{min}}(t)}{\hat{h}_{\mathrm{max}}(t)
-\hat{h}_{\mathrm{min}}(t)}\right)\right] \nonumber \\
\hat{h}(\bbox{r},t)&=&h_{+1}(\bbox{r},t)+h_{-1}(\bbox{r},t)
\end{eqnarray}
This means that the highest actual value, $\hat{h}_{\mathrm{max}}(t)$, always
refers to \emph{black} ($c=0$), whereas the actual minimum value,
$\hat{h}_{\mathrm{min}}(t)$ encodes \emph{white} ($c=255$). Both extreme values
change of course in time, therefore each snapshot of the time series
presented has its own value mapping.

\begin{figure}[htbp]
\centerline{
\epsfig{figure=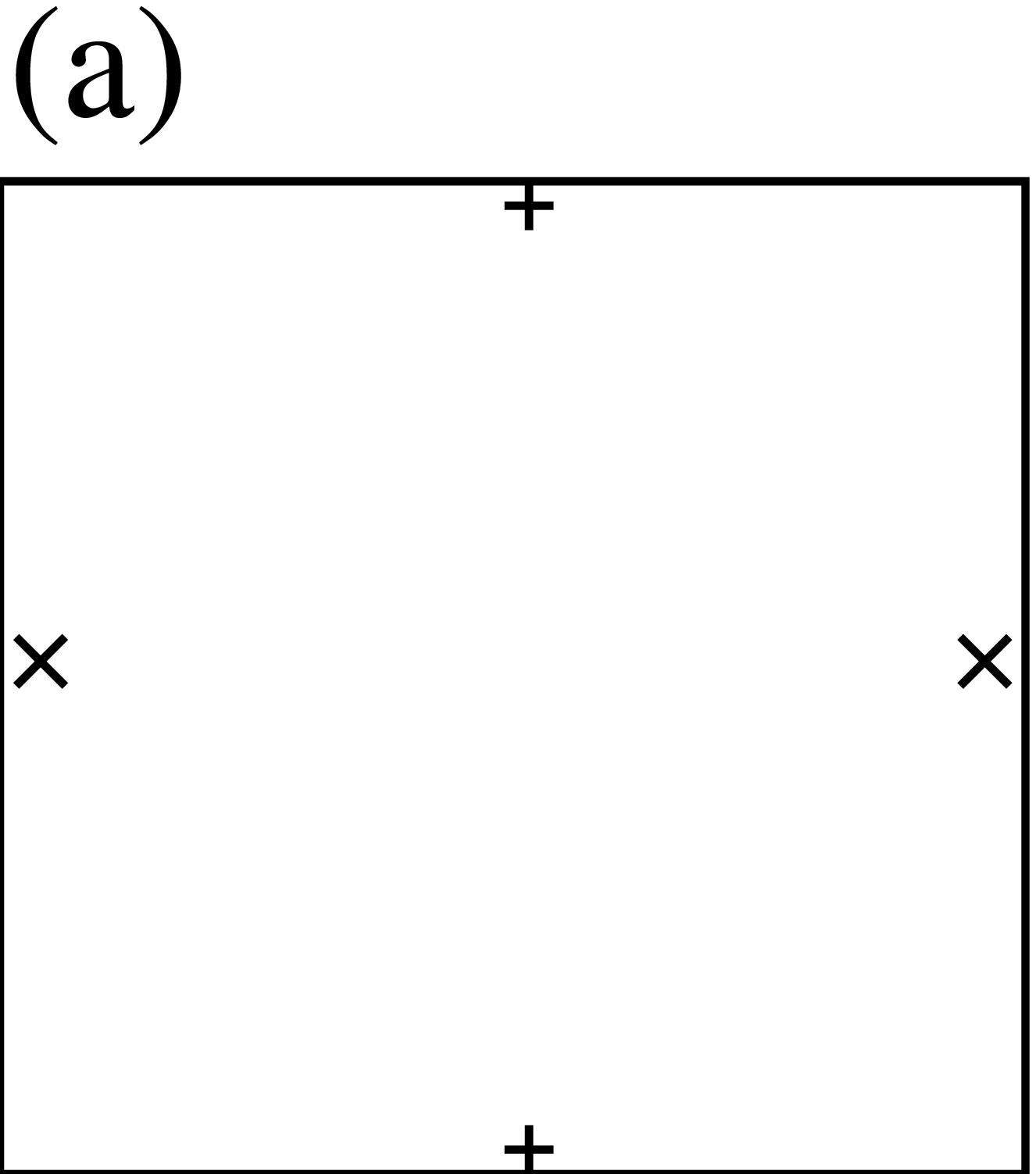,width=2.7cm}
\hfill
\epsfig{figure=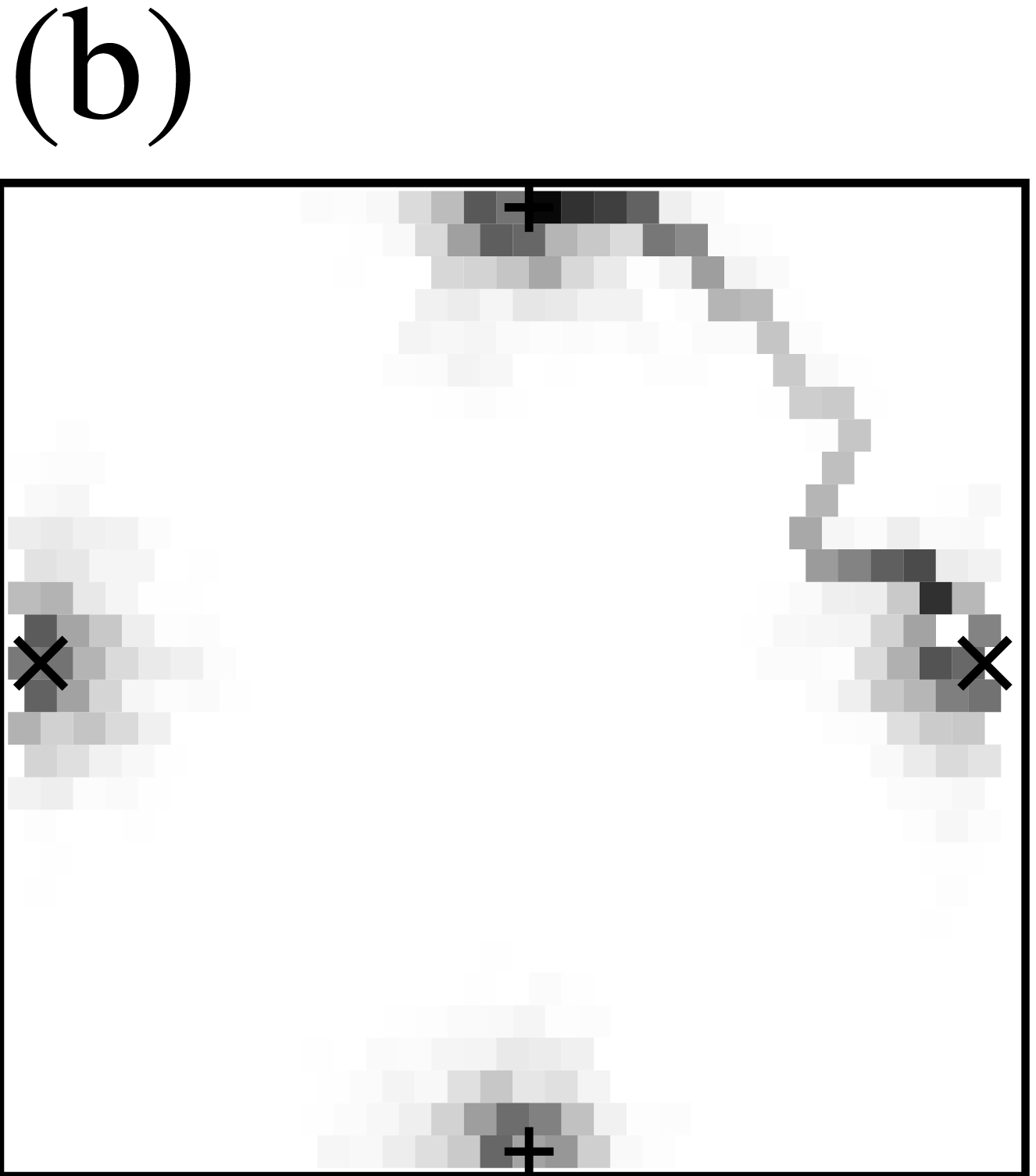,width=2.7cm}
\hfill
\epsfig{figure=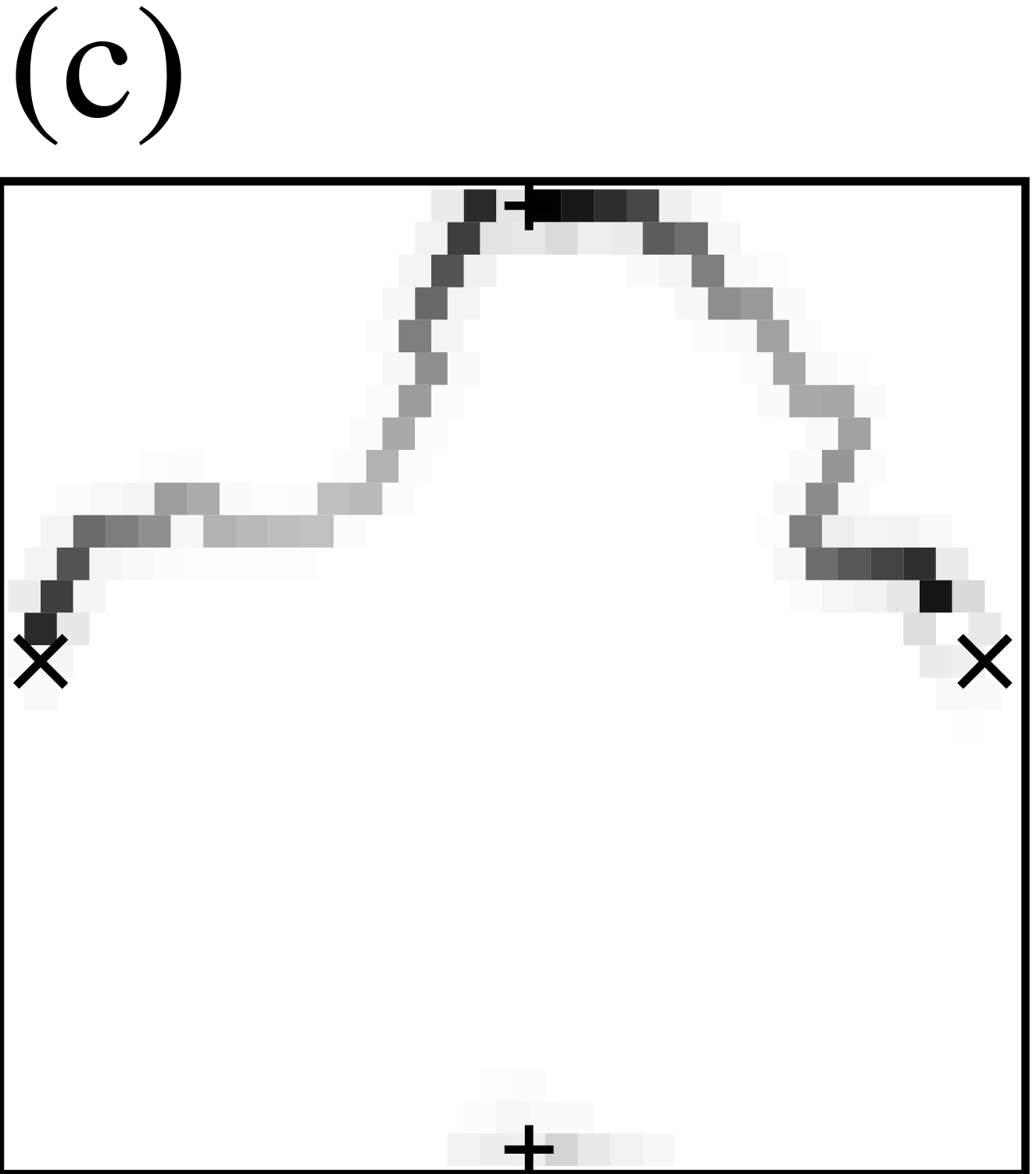,width=2.7cm}
\hfill
\epsfig{figure=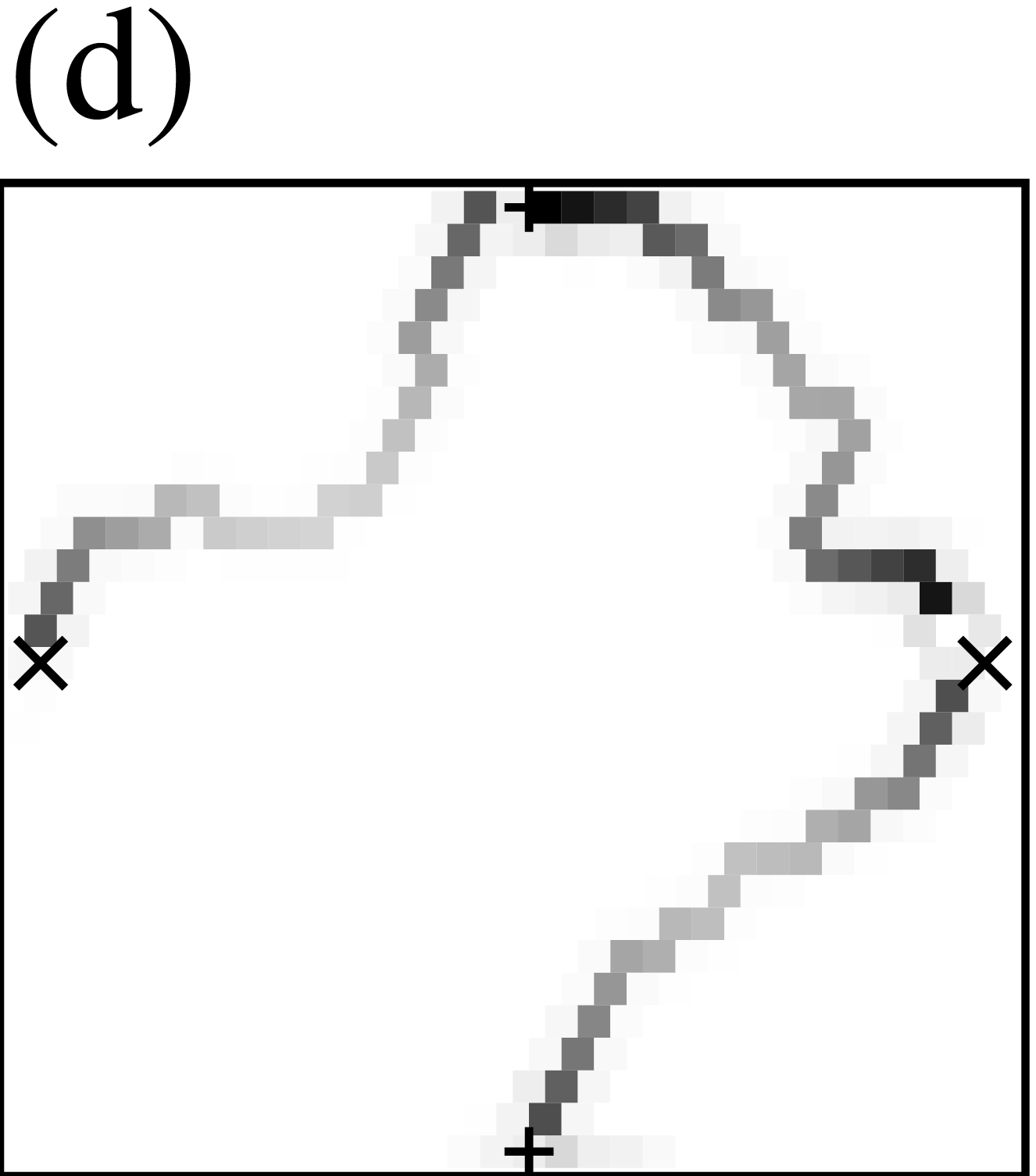,width=2.7cm}
}
\caption[figx]{
  Formation of links between 4 nodes: (a) initial state, (b) after 100,
  (c) after 1.000, (d) after 4.500 simulation steps. Lattice size $30
  \times 30$, 450 agents.  Parameters: $s_{\mathrm{max}}$ = 25.000,
  $k_{h}$ = 0.01, $\beta$ = 0.2, $k_{h}$ =
0.01. 
\label{karo}\citep{fs-tilch-02-a}}
\end{figure}
\begin{figure}[htbp]
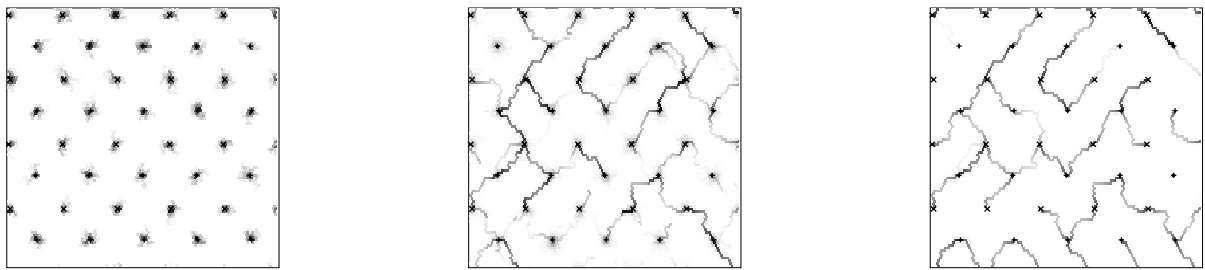

\centerline{
\epsfig{figure=figures/r10.eps,height=3.6cm,width=3.6cm}
\hfill
\epsfig{figure=figures/r100.eps,height=3.6cm,width=3.6cm}
\hfill
\epsfig{figure=figures/r1000.eps,height=3.6cm,width=3.6cm}
}
\caption{
  Time series of the evolution of a network after 10, 100, 1.000
  simulation steps.  Lattice size $100 \times 100$, $5.000$ agents, $40$
  nodes, $z_{+}=20$, $z_{-}=20$.  Parameters: $s_{\mathrm{max}} =
  10.000$, $k_{h}= 0.03$,
  $\beta = 0.2$. 
\citep{fs-tilch-02-a}
\label{t-rautpic}}
\end{figure}

As a first example, we show the evolution of the connections between four
nodes.
Fig.~\ref{karo} would suggest that in the course of time all nodes with
an opposite potential should be connected. This however is not true
because the existing connections cause a \emph{screening effect}, which
forces the agents to move along existing connections rather than making
up new ones. This screening effect becomes more obvious, when the number
of nodes is increased. Fig.~\ref{t-rautpic} shows the time evolution of a
network which should connect 40 nodes, as shown in the setup of
Fig.~\ref{raut0}.
Here, we see that in the course
of time the agents aggregate along the connections, which results in
higher agent concentrations and in higher fields along the
connections.\fn{A video of these computer simulations can be found at
  \url{http://intern.sg.ethz.ch/fschweitzer/until2005/network.html}.
}

\begin{figure}[htbp]
\centerline{\epsfig{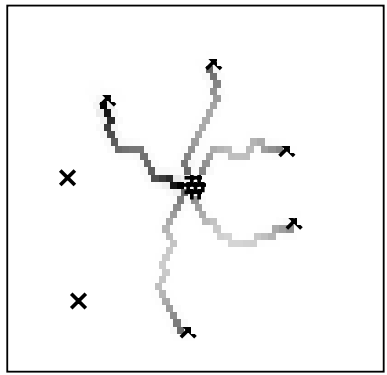}}
\caption{
  Formation of links between a center ($z_{-}=1$) and surrounding nodes
  ($z_{+}=7$) after $10.000$ simulation steps. Lattice size: $50 \times
  50$, $2.000$ agents. Parameters: $s_{\mathrm{max}}=20.000$, $k_{h}=0.02$,
  $\beta = 0.2$. 
\label{cono} \citep{fs-tilch-02-a}}
\end{figure}

In Figs. \ref{karo},\ref{t-rautpic}, the connections between the nodes
exist as a two-component chemical field generated by the Brownian agents.
This self-assembling network is created very fast and remains stable
after the initial period.  We note that the network formation is not
restricted to regular or symmetric distributions of nodes.
Fig.~\ref{cono} shows a simulation, where different nodes are connected
with a center.  An extension of the model has been applied to simulate
the trunk trail formation in ants connecting a nest to different food
sources \citep{fs-lao-family-97} and will be discussed in Secs. 6, 7.

Patterns like the networks shown in Figs.
\ref{karo},\ref{t-rautpic},\ref{cono} are intrinsically determined by the
history of their creation. It means that irreversibility and early
symmetry breaks play a considerable role in the determination of the
final structure. These structures are unique due to their evolution, and
therefore can be understood only within a stochastic approach, as
provided for example by the Brownian agent approach.

\section{Estimation of the Network Connectivity}
\label{sec:estim-netw-conn}

In order to characterize a network, one of the most important questions
is, whether two nodes $k$ and $l$ are connected or not. In the model
considered, a connection is defined in terms of the chemical field
$\hat{h}(\bbox{r},t)$ produced by the agents. In the beginning, however, the
agents randomly visited almost every lattice site before their motion
turned into a bound motion between the nodes. Therefore, the field
$\hat{h}(\bbox{r},t)$ has a non-zero value for almost every $\bbox{r}$, which
exponentially decays, but never vanishes. Hence, in order to define a
connection in terms of $\hat{h}(\bbox{r},t)$, we have to introduce a
\emph{threshold value} $h_{\mathrm{thr}}$, which is the minimum value considered
for a link. More precisely, a \emph{connection} between two nodes $k$ and
$l$ should only exist, if there is a path $s \in A$ between $k$ and $l$
on which the actual value of the field is larger than the treshold value:
\begin{equation}
  \label{thresh}
 \hat{h}(s,t)>h_{\mathrm{thr}} \;\;\; \mbox{for $s \in A$} 
\end{equation}
Such a definition does not necessarely assumes that the connection has to
be a \emph{direct link}. Instead, it could be any path $s$, which may
also include other nodes, as long as the value $\hat{h}(s,t)$ along the
path is above the threshold.

We want to define the \emph{local connectivity} $E_{lk}$ as follows:
\begin{equation}
  \label{connect}
 E_{lk} =   \left \{
 \begin{array}{ll}
1 &  \mbox{if $k$ and $l$ are connected by a path $s \in A$,} \\
 & \mbox{along which $\hat{h}(s,t)>h_{\mathrm{thr}}$}  \\
0 & \mbox{otherwise}
\end{array} \right. 
\end{equation}
We note that the connectivity $E_{lk}$ does not change, if two nodes $k$
and $l$ are connected by more than one path.

If we consider a number of $z$ nodes, then the \emph{global connectivity}
$E$ which refers to the whole network, is defined as follows:
\begin{equation}
  \label{connec-e}
  E=\frac{\sum\limits_{k=1}^{z}\,\sum\limits_{l>k}^{z}\, E_{lk}}{
\sum\limits_{k=1}^{z}\,\sum\limits_{l>k}^{z}\,1}\;=\;
\frac{2}{z(z-1)}\,\sum\limits_{k=1}^{z}\,\sum\limits_{l>k}^{z}\, E_{lk}
\end{equation}
Dependent on the configuration of nodes, there may be numerous different
realizations for the connections, which result in the same connectivity
$E$. 

In order to use the definition for the connectivity to evaluate the
simulated networks, we first have to define the threshold value
$h_{\mathrm{thr}}$. This should be the \emph{minimum value} of $\hat{h}(r,t)$
along a \emph{stable connection} between two nodes. Based on this
definition, we have derived in a recent paper \cite{fs-tilch-02-a}
an estimate for $h_{\mathrm{thr}}$:
\begin{equation}
  \label{tres-f}
  h_{\mathrm{thr}}= \bar{n} \frac{s_{\mathrm{max}}}{k_{h}}\,
\left(\frac{s_{\mathrm{min}}}{s_{\mathrm{max}}}\right)^{1/4}
\end{equation}
Provided the set of parameters, used for the simulations, we find for the
threshold the value $h_{\mathrm{thr}}= 1.7 \times 10^{4}$, which is approximately
$h_{\mathrm{thr}}\approx 2 s_{\mathrm{max}}$. We note, that this is an estimate, which might
give a rather high value and thus ensures, that values for $\hat{h}(s,t)$
above the threshold \emph{really} represent a \emph{stable} connection
$s$.

In a recent paper \cite{fs-tilch-02-a}, we have investigated how
the global connectivity $E$ evolves in the course of time and how it
changes in dependence on the density of agents, $\bar{n}$.  In
agreement with the visible evolution of the network presented in
Fig.~\ref{t-rautpic}, three different stages can be clearly distinguished: (i)
an \emph{initial} period ($t<10^{2}$), where no connections yet exist,
(ii) a \emph{transient} period ($10^{2}<t<10^{4}$), where the network
establishes, (iii) a \emph{saturation} period ($t>10^{4}$), where almost
all nodes are connected, and only small fluctuations in the connectivity
occur.

Here, we focus on the question, how the connectivity changes in
dependence on stochastic influences. According to Eq.~(\ref{u-ik-d}), in
the Brownian agent approach the different state variables of the agent
are changed both by deterministic and stochastic influences.  In our
agent-based model of network formation, stochasticity influences are
explicitely considered only in the equation of motion of the Brownian
agents, Eq. (\ref{langev-red2}). There, it was assumed that the strength
of the stochastic force $\epsilon_{i}$ could be in general an individual
parameter. This idea will be used in Sec. 6, here we want to assume for
the moment that $\epsilon_{i}=\epsilon$ is equal for all agents and can
be further related to the \emph{temperature} $T$, which is a physical
measure of fluctuations in the system, independent of the agents. It was
Einstein, who derived in 1905 a relation between the temperature and the
diffusion of simple Brownian particles, $D=(k_{B}/\gamma_{0})T$, where
the proportionality is given by the Boltzmann constant $k_{B}$
and the friction coefficient $\gamma_{0}$. Identifying $\epsilon=D$,
which is correct for the physical interpretation of the overdamped
Langevin Eq.~(\ref{langev-red2}), we can now express the strength of the
stochastic force by the temperature $T$, which as a global parameter acts
as a boundary condition for the dynamics of the multi-agent system.

\begin{figure}[ht]
\label{e-p}
\centerline{\epsfig{figure=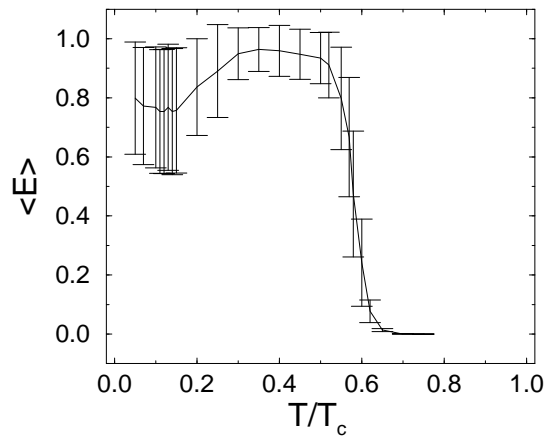,width=7cm}}
\caption[]{
  Network connectivity $E$, Eq.~(\ref{connec-e}), averaged over $200$
  simulations vs.  reduced temperature $T/T_{c}$ Eq.(~\ref{t-crit}). For the
  setup see Fig.~\ref{t-rautpic}. \citep{fsbook03} }
\end{figure}

\begin{figure}[htbp]
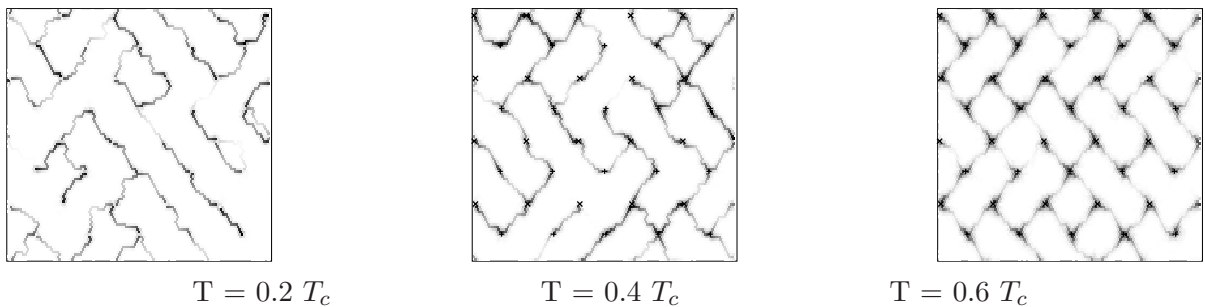

  \centerline{
    \epsfig{figure=figures/netz20.eps,height=3.5cm,width=3.5cm}\hfill
    \epsfig{figure=figures/netz40.eps,height=3.5cm,width=3.5cm}\hfill
    \epsfig{figure=figures/netz60.eps,height=3.5cm,width=3.5cm} }
  \centerline{ T = 0.2 $T_{c}$ \hspace{2.5cm} T = 0.4 $T_{c}$
    \hspace{2.5cm} T = 0.6 $T_{c}$} 
\caption{
  Network formation dependent on the reduced temperature $T/T_{c}$,
  Eq.~(\ref{t-crit}).  Shown is the value $c(\bbox{r},t)$,
  Eq.~(\ref{grey}) after $5.000$ simulation steps.  Other parameters see
  Fig.~\ref{t-rautpic}.
\label{noise-p} \citep{fsbook03}}
\end{figure}

From the physics perspective, network formation -- as other processes of
dissipative structure formation in non-equilibrium systems \citep{biebrep-95} --
should become possible only below a certain critical temperature $T_{c}$.
Above $T_{c}$, fluctuations in the system rather destroy the structures
emerging. With respect to Eq.~(\ref{langev-red2}), this means that for
$T>T_{c}$, the agents cannot pay sufficient attention to the determinstic
influences given by $f_{i}$ and thus behave more or less as simple
Brownian particles. The delicate balance between the stochastic and
deterministic influences of course depend on the current value of
$f_{i}$, which in turn results from the history of the indirect
communication process described by Eqs.~(\ref{h-net-nd}),
(\ref{dh-eff}). However, in a so-called mean-field approximation for the
agent density and the dynamics of the field components \citep{fsbook03} we were
able to derive an expression for the critical temperature $T_{c}$ below
which the network formation becomes feasible: 
\begin{equation}
  \label{t-crit}
  T_{c}=\frac{\alpha}{2} \frac{\bar{s}\, \bar{n}}{k_{B}\,k_{h}}
\end{equation}
This equation relates $T_{c}$ to the mean density of agents,
$\bar{n}=N/A$, and an average production rate of the chemicals $\bar{s}$,
derived in \citep{fsbook03} as:
\begin{equation}
  \label{q-mean2}
  \bar{s}=\frac{s_{\mathrm{max}}\,-\,s_{\mathrm{max}}}{\ln s_{\mathrm{max}}\,-\,\ln s_{\mathrm{max}}}
\end{equation}
By varying the stochasticity, expressed by the parameter $T$ relative to
the reference value $T_{c}$, we are now able to investigate the
connectivity by means of computer simulations, carried out for the same
setup as used for Fig.~\ref{t-rautpic}.  Fig.~\ref{e-p} shows the
connectivity, averaged over 200 simulations, as a function of the reduced
temperature $T/T_{c}$, whereas Fig.~\ref{noise-p} shows some simulated
realizations of the networks for different reduced temperatures.

Fig.~\ref{e-p} clearly indicates that an optimal range of temperatures, $0.3
\leq T/T_{c} \leq 0.5$, exists where the average connectivity reaches a
maximum. For the given setup of nodes and the given set of parameters,
within this optimal range in the average more than 90 percent of the
nodes are connected. For $T>0.5\;T_{c}$, the connectivity breaks down
drastically, which means that the motion of the agents is mainly
determined by the stochastic forces, and no attention to the gradient of
the field is payed. 

On the other hand, for $T<0.3\;T_{c}$, the connectivity decreases because
during the transient period of establishing the network, the agents have
payed too much attention to the gradient, and therefore are trapped in
the first established links, instead of moving around. In this range of
temperatures, the average connectivity $\mean{E}$, Fig.~\ref{e-p},
displays large fluctuations, so almost every realization between $E=0.5$
and $E=1$ has a certain likelyhood.  Here, the early history in creating
the network plays an important role for the eventual pattern and the
connectivity. However, in the optimal range of temperature, the effect of
the early symmetry breaks is weakened to a certain extent, thus leading
to smaller fluctuations in $E$.

We also want to note, that in the range of smaller temperatures, of
course less connections between the nodes occur, but these connections
provide a much stronger link in terms of the field $\hat{h}(\bbox{r},t)$.
One of the reasons for the breakdown of the connectivity for $T>0.5
T_{c}$ in Fig.~\ref{e-p} results from the fact, that many of the links
established have a value below the threshold, $h_{\mathrm{thr}}$, and
therefore are not considered as stable connections. This can be seen by
comparing the pictures in Fig.~\ref{noise-p}. Here, the nodes still seem to
be connected for $T=0.6\;T_{c}$, the value of the related field however
is below the threshold, $h_{\mathrm{thr}}$.

\section{Modeling Trail Formation in Ants}
\label{5.3.2}

In this section, the network model introduced above is extended to
simulate the formation of directed trails, as observed in group-raiding
ants. So, the nodes are given now by a nest and different food sources,
and the links are given by the directed trails between nest and food.
These connections do not exist from the very beginning; they have to be
generated by the agent community in a process of self-organization.
Hence, the agents have to perform two quite different tasks, which are
referred to each other: first, they have to detect food places - unknown
to them, and then they have to link these places to their original
starting point by forming a trail, with local chemical orientation as the
only tool provided. 

For our model, we assume that initially all agents are concentrated in
one place, the ``nest''.  The internal parameter, $\theta_{i}=\pm1$,
describes whether an agent has found food yet or not. If the agent starts
from the nest, the internal parameter is always set to $\theta_{i}=+1$.
Only if the agent has detected a food source, $\theta_{i}$ is changed to
$\theta_{i}=-1$. Dependent on the value of $\theta_{i}$, the agent is
able to produce one of two different chemicals with a rate
$s_i(\theta_i,t)$, which was already described in Eq.~(\ref{prod}).
I.e., the agent produces chemical of component $(+1)$, when it leaves the
nest ($\theta_{i}=+1$) at time $t_{n}^{i}$, and starts to produce
chemical of component $(-1)$ instead, after it has found food
($\theta_{i}=-1$) at time $t_{f}^{i}$.  The dynamics of two chemical
components is again given by Eq.~(\ref{h-net-nd}).  Further, the agents
are affected by the different kind of chemicals as described by
Eqs.~(\ref{langev-red2})(\ref{dh-eff}).

On could argue that for the simulation of foraging trails of ants, the
most simple assumption could be that {\em only} individuals which found
food create a pheromone trail, while turning straight back to the nest.
These trails could then be used by other ants to find the food sources,
once they found the trails. However, those simulations have to overcome
the problem where the first ant which found food got the information from
to find its way back home. In biological systems, geocentric or
egocentric navigation could provide this additional information needed
for successfully returning to the nest.  Since our model simulates the
formation of trunk trails without counting on navigation and internal
storage of information, our main assumption is that a simple chemotactic
response would be sufficient to generate trails between the nest and the
food sources. However, we have to distinguish trails which have been
created during the (unsuccessful) search for food from trails which
really lead to a food source. In our model, this additional information
is encoded by assuming that the agents produce, and respond to, two
different chemical fields.

For the computer simulations, we have to address an additional problem
that results from obvious differences in complexity between biological
entities and Brownian agents.  It is known from central foragers like
ants that they are able to leave a place where they don't find food and
reach out for other areas. This indicates that they have at least a
certain ability to increase their mobility since they have a certain aim,
finding food. Agents in our model however do not have aims and do not
reflect a situation. So, they stick on their local markings even if they
they did not find any food source. In order to increase their mobility in
those cases, we assume that every agent is described by an additional
internal variable, its individual sensitivity $\eta_{i}$ to follow a
chemical gradient.  As long as the agent does not hit a food source, this
sensitivity is continuously decreasing via an increase of the
stochasticity $\eps_{i}$ in Eq.~(\ref{langev-red2}), i.e. we assume here
$\eta_{i}=\alpha/\epsilon_{i}$, where $\alpha$ is the response
parameter to weight the influence of the gradients. Increasing the
individual stochasticity then means that the agent more and more ignores
the chemical detected and thus becomes able to choose also sites not
visited so far.

However, if the agent does not find any food source after a certain
number of steps, the sensitivity $\eta_{i}$ becomes negligible, and the
agent only behaves like a Brownian particle. From the utilitarian
perspective, such a agent becomes ``useless'' for the exploitation of
food sources, if its individual noise level, $\eps_{i}$, exceeds a
maximum value, $\eps_{\mathrm{max}}$ (which could be related to the
critical temperature, discussed above). So we simply assume that the
agent ``dies'' at this level and is removed from the system.  On the
other hand, if the agent hits a food source, its sensitivity is set back
to the initial high value and is kept constant to increase the chance
that the agent finds its way back along the gradient. The switch of the
level of sensitivity is accompanied with the use of the two different
chemical markers, and in this particular sense reminds on
\emph{chemokinesis} \citep{dunn-83}, where the level of activity can be
changed due to chemical conditions in the surrounding.

Using the internal parameter $\theta_{i}$ again, we may assume that the
sensitivity $\eta_{i}$ of the agent changes according to: 
\begin{equation}
\frac{\alpha}{\eta_{i}(t)}=\epsilon_{i}(t)=\frac{\theta_i}{2}\,\left\{\;
(1+\theta_i)\,\Big[\eps_{0}+\delta\eps\,(t-t_{n}^{i})\Big] \;-\;
(1-\theta_i)\,\eps_{0}\;\right\}
\label{rec-noise}
\end{equation}
Here, $\delta\eps$ is the increase of the individual noise level,
$\eps_{i}$, per time step if the agent has not found any food source.

This way, we have completed our model for the trail formation in ants.
The basic ingredient is the chemotactic response to a chemical gradient
which has been produced by the agents themselves.  Hence, the basic
equations are given by the equation of motion for the agents,
Eq.~(\ref{langev-red2}), which is coupled with the equations for the two
chemical fields, Eq.~(\ref{h-net-nd}).  We note that, with respect to
biology, there are different parameters which may influence trail
following in addition to sensitivity, such as trail fidelity, traffic
density, detection distance, endurance of the trail, navigation
capabilities etc.  \citep{haefner-crist-94,edelstein-keshet-et-95}.
Contrary, our model considers only minimal assumptions for the trail
formation.  Here, the formation of trail patterns is solely based on
simple \emph{local} chemical communication between the agents, with no
additional capabilities of orientation or navigation.  The computer
simulations outlined below should prove that the simple local rules
assumed for the action of the agents are sufficient enough to solve such
a complex problem.

\section{Computer Simulations of Trail Formation in Ants}
\label{sec:ants}

Different from arbitrary {\em non-directed} tracks which are commonly
used by the agents but lead to nowhere, {\em directed trails} should link
a starting point (e.g. a nest) to a destination point (e.g. a food
source) and back.  In order to define start and destination, we assume on
the two-dimensional surface a center (nest) where the agents are
initially concentrated, and random distribution of five separated food
sources food sources unknown to the agents.  As in the network model
described before, both the nest and the food sources, {\em do not
  attract} the agents by a certain long-range attraction potential, they
are just particular sites of a certain size (defined in lattice sites).
It is assumed that the food sources can be exhausted by the visting
agents, which are assumed to carry part of them back to the nest.

\begin{figure}[htbp]
\centerline{\epsfig{figure=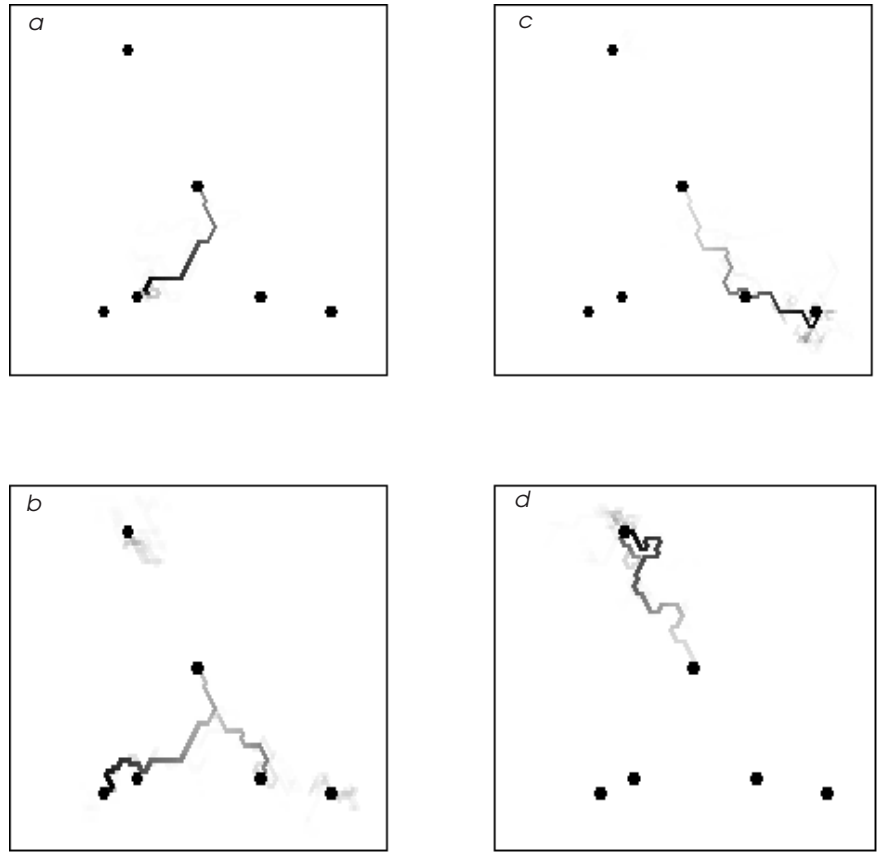,width=11.5cm}}
\caption[fig4]{
  Formation of trails from a nest (middle) to five randomly placed food
  clusters. The distribution of chemical component $(-1)$ (see text) is
  shown after (a) 2000, (b) 4000, (c) 8500, and (d) 15000 simulation time
  steps, respectively. \citep{fs-lao-family-97}
  \label{5food-sim}}
\end{figure}

Fig.~\ref{5food-sim} shows a time series of the  trail formation in terms
of the spatial concentration of chemical component $(-1)$, which is coded
again in a grey scale.  In Fig.~\ref{5food-sim}a, we see that two of the food
sources randomly placed on the lattice, have about the same distance to
the nest, but the one, which - by chance - has been discovered first,
will also be the one linked first to the nest. This reflects the
influence of \emph{initial symmetry breaking effects}.

If the food sources are exhausted by the agents carrying food to the
nest, they vanish. Therefore, the individual sensitivity of the agents
coming from the nest to the food cannot be set back to the high initial
value and the agents further increase their mobility by ignoring the
trail, they reach out again, and by chance discover new food sources. But
since a trail already exists, those sources in a close vicinity to the
one that disappeared have a larger probability to be discovered. Part of
the ``old'' trail is re-used then to link the new food source to the nest
(compare Fig.~\ref{5food-sim}a,b,c).

Fig.~\ref{5food-sim}b shows that at the same time also more than one
sources could be linked to the nest by trails. However, all trails
compete for the agents to be maintained, and the trails could survive
only if the concentration of the chemicals is above a certain critical
value.  Therefore, the coexistence of different trails only lasts for a
short time. Hence, the trails to the different food sources will appear
one after the other, with the old trails disappearing again by
decomposition because they are no longer maintained.  After a source is
exhausted, we usually observe a time lag before the next one is
exploited. This is needed for the system to ``forget'' about the useless
trail and to establish a new trail to the next source.  Thus, we can
distuingish between two alternating dynamical stages: (i) a stage of
rather random motion of the agents (\emph{exploration} of the sources),
and (ii) a stage of directed motion of the agents along the trail
(\emph{exploitation} of the sources).

The flexibility of the model is indicated by the fact, that even after a
long period of trail formation in the lower part of the lattice, the
agents are able to link the food source on the upper left side of the
lattice to the nest (Fig.~\ref{5food-sim}d) and therefore finally have
detected and linked all five sources to the nest.  The desorienation
stage before exploiting the last source is especially long, because the
agents could not build on any previous part of a trail into that
direction, whereas the desorientation stage before exploiting the third
and the fourth source is very short, because of the trails which already
exist into that area.

\section{Conclusions}
\label{sec:concl}

In this chapter, we have proposed a multi-agent approach to model the
self-organization of networks. This approach is based on the interactions
of the agents on a local or ``microscopic'' level, which could lead to
the emergence of the structure as a whole on the global or
``macroscopic'' level.  In our model, the \emph{emergence} of network
structures, i.e., the \emph{self-organized} formation of links results
from the dynamic interaction of the agents, which is different from
drawing lines between a set of known nodes. The agents have to solve two
different tasks in parallel, namely to discover new nodes and to link
them to the network \emph{without} external guidance.  They also have to
ensure the adaptivity of the network.

In our model, the self-assembling of networks between arbitrary nodes is
based on the nonlinear interaction of Brownian agents.  Different from a
circuit diagram, for instance, which determines the different links
between the nodes in a \emph{top-down} approach of hierarchical planning,
the connections here are created by the agents \emph{bottom-up}.  The
spontaneous emergence of different network structures was characterized
as a self-organizing process.  As in every evolutionary game, also for
the emergence of networks critical parameters exist. As an example, we
have discussed the strength of fluctuations compared to the influence of
the chemical gradients, but this also holds for other parameters in the
model.  If, for instance, the decay rate for the decomposition of the
chemicals is too high, or the sensitivity of the agents is too low, they
loose their local orientation and get lost, so that no links appear. If,
on the other hand, the decay rate is too low, it takes a much longer
time, before distinct links appear, because the selection pressure is too
low.

The model turned out to be very flexible regarding the geometry of the
nodes to be connected.  The locations of the different nodes act as a
boundary condition for the structure formation, but do not determine the
way of connecting the different nodes. If, for example, a particular link
breaks down, the agents would be able to \emph{repair} it by
re-establishing the field, again, or by creating a new one. It was shown
\citep{fsbook03} that the switching behavior between a connected and a
disconneced state can be very short, which would allow the construction
of a dynamic switch.

To maintain a network, to keep it updated, to repair links, are ambitious
aims for which natural paradigms exist, e.g. in social insects or neural
networks. These have also inspired the design of our multi-agent model.
Compared to the complex ``individual-based'' models in ecology
\citep{huston-et-88,deanged-92} the agent model proposed here provides a
very simple but efficient tool to simulate a specific structure with only
a view adjustable parameters.  The major difference to biology is denoted
by the fact, that the agents used in the simulations do not have
individual capabilities to store and process informations about the
locations of the nodes, or to count on navigational systems or additional
guidance. They rather behave like physical particles which respond to
local forces in a quite simple manner, without ``implicit and explicit
intelligence'' \citep{haefner-crist-94}. Noteworthy, these local forces
only result from the chemical markings produced by themselves.  This way,
we propose a kind of minimalistic approach for the formation of networks
which is build on the dynamical essence of the feedback processes rather
than on the internal complexity of the agents.

\bibliography{web-fs-network}

\end{document}